\documentclass[twocolumn,prl,superscriptaddress,showpacs,letterpaper]{revtex4}
\usepackage{graphicx,amsmath,amssymb,dcolumn}

\newcommand{\F}{\mathcal{F}}

\begin{document}
\title{Reaching optimal efficiencies using nano-sized photo-electric devices}
\author{B. Rutten}
\affiliation{Hasselt University - B-3590 Diepenbeek, Belgium}
\author{M. Esposito}
\altaffiliation{Also at Center for Nonlinear Phenomena and Complex Systems, Universit\'{e} Libre de Bruxelles, Code Postal 231, Campus Plaine, B-1050 Brussels, Belgium.}
\affiliation{Department of Chemistry and Biochemistry and Institute for Nonlinear Science,
University of California, San Diego, La Jolla, CA 92093-0340, USA}
\author{B. Cleuren}
\email{bart.cleuren@uhasselt.be}
\affiliation{Hasselt University - B-3590 Diepenbeek, Belgium} 
 
\begin{abstract}
We study the thermodynamic efficiency of a nano-sized photo-electric device and show that at maximum power output, the efficiency is bounded from above by a result closely related to the Curzon-Ahlborn efficiency. We find that this upper bound can be attained in nano-sized devices displaying strong coupling between the generated electron flux and the incoming photon flux from the sun. 
\end{abstract}
\date{\today}

\maketitle
Understanding and controlling the mechanisms that determine the efficiency of photo-electric devices is of fundamental importance in the quest for efficient and clean sources of energy. Thermodynamically speaking, these devices are driven by the temperature difference between a hot reservoir (sun, temperature $T_{s}$) and a cold reservoir (earth, ambient temperature $T$). Therefore, like any heat engine, the efficiency at which the conversion of radiation into electrical energy takes place has a universal upper bound given by the Carnot efficiency $\eta_{c}=1-T/T_{s}$ \cite{callen}. Although this result has fundamental theoretical implications, it is of poor practical use since it is only reached when the device is operating under reversible conditions. Hence the generated power, defined as the output energy divided by the (infinite) operation time, goes to zero. In realistic circumstances of finite power output, the efficiency will necessarily be below the Carnot limit due to irreversible processes taking place in the device. Another source of possible efficiency decrease are energy losses within the device for example due to non-radiative recombination of charge carriers. Since the operational parameters of the device are mostly determined in such a way that a maximum power output is obtained, Curzon and Ahlborn examined in 1973 the efficiency of a Carnot cycle with a finite cycling time and, using the endoreversible approximation, found an efficiency at maximum power $\eta_{ca}=1-\sqrt{T/T_{s}}$ \cite{ca}. This result is remarkable since it does not depend on the specific details of the system, and thus the question of universality naturally arises. Recent works \cite{vdbPRL,SanchoPRE06,vdbADVCP,universal,QD} have indeed demonstrated that in the linear regime (small temperature differences, $\eta_{c}\ll 1$) the Curzon-Ahlborn efficiency is universal for so-called \emph{strongly coupled systems}, where the heat and work producing fluxes are proportional. In these systems internal energy losses are absent, implying that the resulting efficiency is exclusively determined by the (unavoidable) irreversible processes occurring at finite power. Hence, at least in the linear regime, the Curzon-Ahlborn efficiency is indeed a universal upper bound, with a similar status as the Carnot efficiency. In the non-linear regime, the efficiency at maximum power becomes device dependent but is again found to be highest for strongly coupled systems. Remarkably, it remains closely related to the Curzon-Ahlborn result \cite{universal,QD}.

While energy losses are almost unavoidable in the macroscopic world, new technological developments at the nano-scale open up the road to highly efficient devices. In thermoelectric research it is well established that the use of low-dimensional, nanostructured devices significantly increases the efficiency. Such devices have a sharply peaked density of states, a prerequisite for a good thermoelectric \cite{hicks,mahan,02LinkePRL,05LinkePRL,QD}. A similar tendency towards the development of nanostructured materials and even single nano-sized devices also occurred in photovoltaic applications \cite{schallerPRL2004,klimovAPL2006,timmermanNAT2008,kelzenberg,lieberNAT,lieberREV}. 

In view of these recent developments in nonequilibrium thermodynamics as well as in nanotechnologies, we propose in this letter to investigate the performance of a single nano-sized photo-electric device. The discrete nature of its energy levels is essential to provide strong coupling and thus high efficiencies. A detailed microscopic description of the device dynamics is presented, which allows for an exact analysis of the efficiency, far into the non-linear regime. Our central result is that this device displays strong coupling when non-radiative recombination processes can be ignored. The corresponding efficiency at maximum power is then found to be very close to the Curzon-Ahlborn result. More generally, strong coupling provides a guiding principle in the quest for building high efficiency photo-electric devices. Experimentally, the degree of coupling can be determined by measurement of the Onsager coefficients.

\begin{figure}
\includegraphics[width=2.6in]{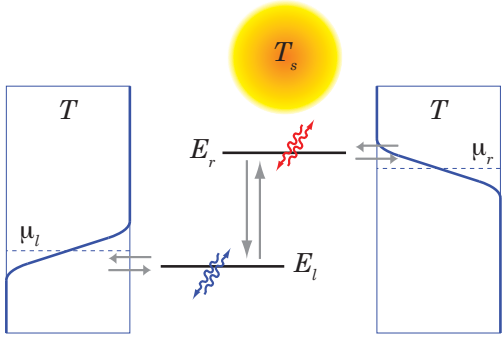}
\caption{(Color online) Schematic view of the nano-sized photo-electric device. The grey arrows show the different allowed electron transitions. Transitions between the two energy levels are induced by solar photons (red curved arrows) and by non-radiative processes (blue curved arrows).}
\label{fig:setup}
\end{figure}

The nano-device we consider is composed of two single particle levels of energy $E_{l}$ and $E_{r}(>E_{l})$, which define the bandgap energy $E_{g}=E_{r}-E_{l}$. We assume that Coulomb interactions prevent two electrons to be present at the same time in the device. As a result, the device is either empty ($0$) or has one electron in level $E_l$ or $E_r$ with respective probabilities $p_{i}$ with $i \in \{0,l,r\}$. The device is connected with two leads ($l$ and $r$). The left (right) lead can only exchange electrons with the level $E_{l}$ ($E_{r}$) as illustrated in Fig.~\ref{fig:setup}. Such a nano-device could be made for example of two coupled single-level quantum dots, each connected to a given lead.
The leads are at the same temperature $T$ but have different chemical potentials $\mu_{l}$ and $\mu_{r}=\mu_{l}+qV$ due to an applied voltage ($q$ is the electron charge). Electron transitions between $E_l$ and $E_r$ are induced by two possible mechanisms. The first is due to the incoming sun (black body) radiation at the resonant energy $h\nu=E_{g}$. The second is due to non-radiative processes at the same resonant transition. The dynamics of the cell is described using a master equation formulation for driven open systems \cite{EspositoHarbola07PRE,universal} presented below. Such a description can be shown to be equivalent to the quantum dynamics systematically derived from the microscopic Hamiltonian in Refs. \cite{GalperinPRL,GalperinJCP} when the level broadening is smaller than $E_{g}$. The master equation reads
\begin{equation}
\left[\!\!
\begin{array}{c}
\dot{p}_{0}(t)\\ \dot{p}_{l}(t)\\ \dot{p}_{r}(t)\\
\end{array}\!\!\!
\right]
\!\!=\!\!
\left[
\begin{array}{ccc}
\!\! -k_{l0}\!-\!k_{r0}                 & k_{0l}                  & k_{0r}\\
k_{l0}			                &\!\!-k_{0l}\!-\!k_{rl}   & k_{rl}\\
k_{r0}			                & k_{rl}		  &\!\!-k_{0r}\!-\!k_{lr} \!\!
\end{array}
\right]\!\!\!\!
\left[\!\!
\begin{array}{c}
p_{0}(t)\\ p_{l}(t)\\ p_{r}(t)\\
\end{array}
\!\!\!\right].
\end{equation}
where $k_{ij}$ denotes the transition rate from state $j$ to $i$. The rates describing the exchange of electrons with the leads are given by
\begin{equation}
\begin{array}{lcll}
k_{l0} = \Gamma_{l}f(x_{l}) & ; &  k_{0l} = \Gamma_{l}[1-f(x_{l})] & ; \\
k_{r0} = \Gamma_{r}f(x_{r}) & ; &  k_{0r} = \Gamma_{r}[1-f(x_{r})] & ,
\end{array}
\end{equation}
where $f(x) = [\exp(x)+1]^{-1}$ is the Fermi distribution. The arguments are the scaled energies $x_{l}=(E_{l}-\mu_{l})/(k_{B}T)$ and $x_{r}=(E_{r} - \mu_{r})/(k_{B}T)$ with $k_B$ is the Boltzmann constant. The rates describing the transitions between energy levels due to non-radiative effects ($nr$) and to sun photons ($s$) are given by
\begin{equation}
\begin{array}{ll}
k_{rl} = \Gamma_{nr}n(x_{g})+\Gamma_{s}n(x_{s}) & ; \\
k_{lr} = \Gamma_{nr}[1+n(x_{g})]+\Gamma_{s}[1+n(x_{s})] & ,    
\end{array}
\end{equation}
where $n(x) = [\exp(x)-1]^{-1}$ is the Bose-Einstein distribution with scaled energies $x_{g}=E_{g}/(k_{B}T)$ and $x_{s}=E_{g}/(k_{B}T_s)$. Notice that the ratio of the forward and backward transition rates associated to a given elementary process satisfies the detailed balance condition. This ensures that the equilibrium distribution (when $\mu_{l}=\mu_{r}$ and $T=T_{s}$) has the corresponding grand-canonical form. The electron current entering the device from the left lead is given by
\begin{eqnarray}
J=k_{l0}p_{0}-k_{0l}p_{l}.
\end{eqnarray}
From now on, we will focus on the steady state dynamics of the device defined by $\dot{p}_{0}(t)=\dot{p}_{l}(t)=\dot{p}_{r}(t)=0$. $J$ becomes the current of electrons through the device (positive from left to the right) with a corresponding electric current $qJ$. It can be decomposed as $J=J_{s}+J_{nr}$ with $J_{s}$ and $J_{nr}$ the contributions to the current due to the interaction with the sun and the non-radiative processes respectively:
\begin{eqnarray}
&J_{s}&=\Gamma_{s}n(x_{s})p_{l}-\Gamma_{s}(1+n(x_{s}))p_{r}\;\;\; ;\\
&J_{nr}&=\Gamma_{nr}n(x_{g})p_{l}-\Gamma_{nr}(1+n(x_{g}))p_{r}.
\end{eqnarray}
From a thermodynamic viewpoint, solar cells are heat engines converting part of the heat input from the hot reservoir (the sun) into work by moving electrons from lower to higher chemical potentials. The remaining heat gets transferred to the colder reservoir (the earth). Since all photons interacting with the solar cell have an energy $E_{g}$, the net heat flux coming from the sun (i.e. the net energy absorbed per unit time) is $\dot{Q}_{s}=E_{g}J_{s}$. The heat flux coming from the cold reservoir has three contributions:  $\dot{Q}_{l}=(E_{l}-\mu_{l})J$ and $\dot{Q}_{r}=-(E_{r}-\mu_{r})J$ are due to electron exchanges between the cell and the left and right lead respectively, and $\dot{Q}_{nr}=E_{g}J_{nr}$ is due to the non-radiative energy exchanges. The power $P$ generated by the solar cell to bring electrons from the left to the right lead is given by
\begin{eqnarray}
P=(\mu_{r}-\mu_{l})J=T_s \lbrack x_s-(1-\eta_c)(x_{r}-x_{l}) \rbrack J. \label{Power}
\end{eqnarray}
We verify that $P=\dot{Q}_{l}+\dot{Q}_{r}+\dot{Q}_{nr}+\dot{Q}_{s}$ since energy inside the cell is conserved at steady state. The efficiency at which this conversion takes place is then
\begin{eqnarray}
\eta&=&\frac{P}{\dot{Q}_{s}}=\frac{(\mu_{r}-\mu_{l})J}{(E_{r}-E_{l})J_{s}} \nonumber\\
&=& \left(1-(1-\eta_c)\frac{x_{r}-x_{l}}{x_{s}}\right)\left(1+\frac{J_{nr}}{J_{s}}\right) \label{Effi}.
\end{eqnarray}
The entropy $S(t)$ of the solar cell can be expressed in the usual form $S(t)=-k_B\sum_{i}p_{i}(t)\ln p_{i}(t)$. Its time evolution can be separated in a reversible and irreversible part, $\dot{S}=\dot{S}_{e}+\dot{S}_{i}$, with $\dot{S}_{e}=\dot{Q}_{s}/T_{s}+(\dot{Q}_{l}+\dot{Q}_{r}+\dot{Q}_{nr})/T$ corresponding to the entropy change due to the heat exchange with the different reservoirs and where $\dot{S}_{i} \geq 0$ can be identified as the internal entropy production due to dynamical processes within the solar cell \cite{schnakenberg,jiuli,universal}. In the stationary regime $\dot{S}=0$ so that $\dot{S}_{i}=-\dot{S}_{e}$ and the entropy production takes on the familiar bilinear form
\begin{equation}
\dot{S}_{i}=\dot{Q}_{s} \F_{U}+ J \F_{N}=(x_r-x_l)J-x_s J_s-x_g J_{nr} \label{EntropyProd}
\end{equation}
where $\F_{U}=1/T-1/T_{s}$ and $\F_{N}=(\mu_{l}-\mu_{r})/T$ are the thermodynamic forces conjugated to the energy and matter fluxes respectively. Rearranging this expression leads to
\begin{equation}
qV=\eta_c E_{g}(J_{s}/J)-T\dot{S}_{i}/J,
\end{equation}
which relates the work $qV$ done by the solar cell by moving a single electron up the potential gradient, to a fraction of the incident photon energy minus the irreversible losses. It reduces to the well known ideal cell formula $qV=E_{g}\eta_c$ when the device operates reversibly ($\dot{S}_{i}=0$) \emph{and} when $J_{s}=J$. This last condition implies that non-radiative recombination processes are absent, i.e. $J_{nr}=0$. Such an ideal situation was also considered by Shockley and Queisser in there seminal paper on the efficiency of $p$-$n$ junction solar energy converters \cite{shockJAP1961}. For our nano solar cell it means that the heat and particle flows are proportional, $\dot{Q}_{s}=E_{g}J$, a condition which is identified as thermodynamical strong coupling \cite{kedem,vdbPRL,universal}. For each electron transferred between the two leads, exactly one photon is involved. As we show below, this condition minimizes the entropy production and yields the maximal possible efficiency.

We start our analysis with a focus on the linear regime, close to thermal equilibrium. In this regime, characterized by small thermodynamic forces, the heat and particle flows appearing in the entropy expression Eq.~(\ref{EntropyProd}) can be expanded to first order:
\begin{equation}
\begin{array}{lcll}
\dot{Q}_{s}&=& L_{UU}\F_{U}+L_{UN}\F_N & ; \\
J&=& L_{NU}\F_{U}+L_{NN}\F_N & .
\end{array}
\end{equation}
The coefficients $L_{ij}$ appearing are the well known Onsager coefficients. The off-diagonal elements are responsible for the energy conversion process, and satisfy the Onsager symmetry $L_{UN}=L_{NU}$. For a given temperature difference (quantified by $\F_{U}$), the power $P=-T\F_{N}J$ is maximal for $\F_{N}=-(L_{NU}/2L_{NN})\F_U$, which is (in the linear regime) exactly half the open circuit voltage (divided by the temperature $T$). The corresponding efficiency, 
\begin{equation}\label{effq}
\eta=\frac{\eta_c}{2}\frac{\kappa^2}{2-\kappa^2},
\end{equation}
is precisely half the Carnot efficiency multiplied by a factor depending on the \emph{coupling parameter} $\kappa=L_{UN}/\sqrt{L_{UU}L_{NN}}$ \cite{kedem,vdbPRL} which has a numerical value between $-1$ and $+1$ since $\dot{S}_{i}$ must always be positive. Here it is given by:
\begin{equation}\label{nanoq}
\kappa^{2}=\frac{e^{x_l}(e^{x_g}-1) \Gamma_l \Gamma_r \Gamma_s}{
\begin{array}{lr}
\Big[\Gamma_{nr}(\Gamma_l+\Gamma_r)+e^{x_l}\big((\Gamma_{nr}-\Gamma_l)\Gamma_r
\\ \;\;\;\;\;\;\;\;\;\; +e^{x_g}\Gamma_l(\Gamma_{nr}+\Gamma_r)\big)\Big](\Gamma_{nr}+\Gamma_s)
\end{array}}.
\end{equation}
As is clear from Eq.~(\ref{effq}), the efficiency is maximal for $\kappa=\pm 1$, corresponding to a \emph{strongly coupled} system. From Eq.~(\ref{nanoq}) this requires $\Gamma_{nr}=0$. Conversely, the entropy production at maximal power,
\begin{equation}
\dot{S}_i=\F_{U}^{2}L_{UU}\left[1-(3/4)\kappa^2\right] ,
\end{equation}
reaches its minimal value in strongly coupled systems. We note that this reasoning is valid for any type of heat conversion device. And so, strong coupling can be used as a guiding principle in the development of highly efficient devices. As we demonstrate here, the use of nano-sized devices provides an elegant solution to achieve this in practice.
\begin{figure}[!t]
\includegraphics[width=3.2in]{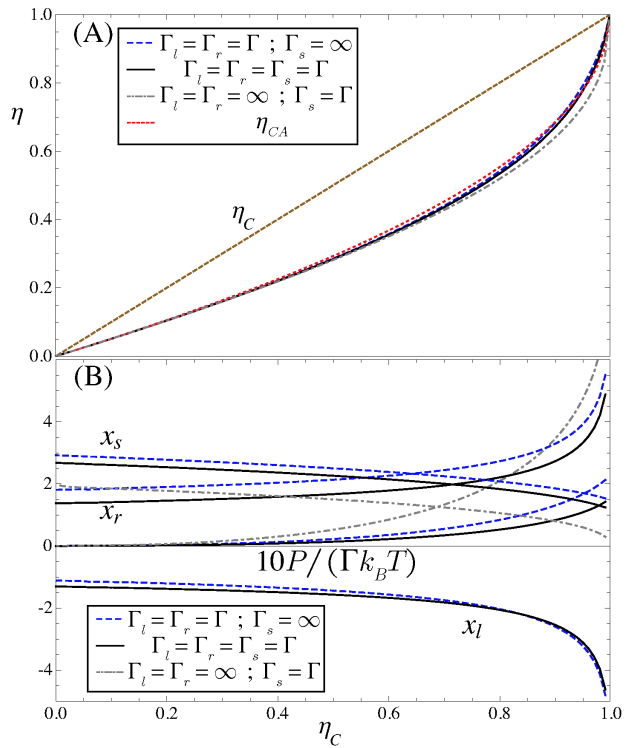}
\caption{(Color online) a) The efficiency at maximum power in a strongly coupled device ($\Gamma_{nr}=0$) as a function of $\eta_c$. b) The corresponding values of the scaled energies $x_l$, $x_r$ and $x_s$. Results are given for different values of the coupling constants $\Gamma_l$, $\Gamma_r$ and $\Gamma_s$.}
\label{fig:strong}
\end{figure}

\begin{figure}[!t]
\includegraphics[width=3.2in]{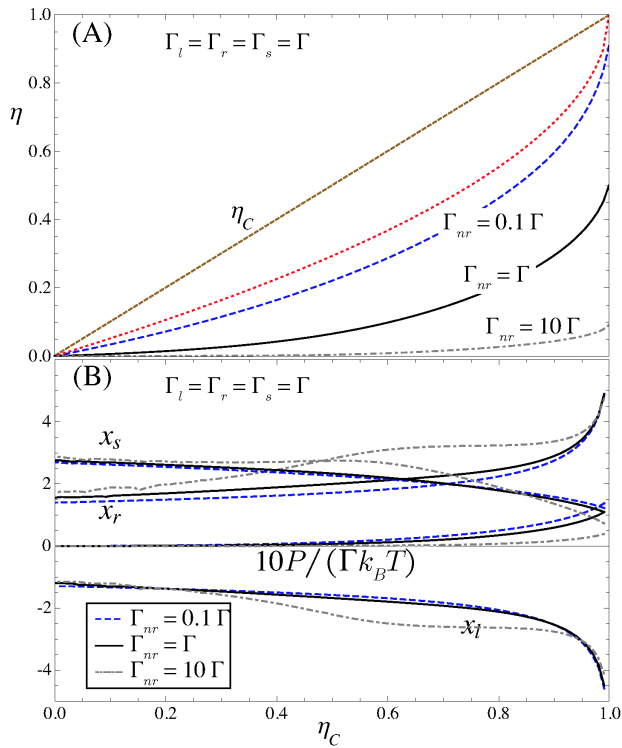}
\caption{(Color online) Same legend as Fig.~\ref{fig:strong} but for a device where non-radiative effects break down the strong coupling condition.}
\label{fig:nostrong}
\end{figure}

We now extend our analysis to the non-linear regime by using the full fledged analytical expressions for the various fluxes. For given values of $\eta_c$ and of $\Gamma$'s, the maximum power output with respect to $x_r$, $x_l$ and $x_s$ cannot be found analytically. However, in the strong coupling case where $\Gamma_{nr}=0$ the numerical search for the maximum can be improved by carrying out some partial maximization analytically. Indeed, by defining $y=x_r-x_l$ and using (\ref{Power}), the condition $\partial_{y}P=0$ together with $\partial_{x_s}P=0$ implies the relation $\partial_{y} J=-(1-\eta_c)\partial_{x_s}J$ which can be solved analytically for $y$ as a function of $x_s$ (the cumbersome expression is not given here but can be derived using simple algebra). Using this solution in the expression for power implies that maximum power can be obtained numerically by finding the maximum with respect to the two remaining variables $x_s$ and $x_l$. In the strong coupling regime, the results of the optimization are shown in Fig.~\ref{fig:strong}. The efficiency at maximum power is plotted as a function of $\eta_c$ for different sets of $\Gamma$'s, together with the corresponding values for $x_l$, $x_r$ and $x_s$ that maximize the power. The efficiency remains remarkably close to the Curzon-Ahlborn result for almost all values of $\eta_c$ and for the different sets of $\Gamma$'s. Slight deviations are only observed far from equilibrium when $\eta_c$ is large. When, due to the presence of non-radiative effects, the strong coupling condition is lost, the results in Fig.~\ref{fig:nostrong} show that the efficiency at maximum power is dramatically decreased below the Curzon-Ahlborn result. Setting $\Gamma_l=\Gamma_r=\Gamma_s=\Gamma$, Table~\ref{table} summarizes the results obtained under practical conditions, i.e. by setting $T=300$K and $T_s=5780$K corresponding to $\eta_c \approx 95\%$. An efficiency of 77.5\% is obtained in the strong coupling case, which is slightly below the Curzon-Ahlborn result ($\eta_{ca}\approx 77.6\%$). Orders of magnitude remain the same when changing the relative values between the $\Gamma$'s

In summary, using a stochastic thermodynamics description of photo-electric devices, we have provided evidence that best efficiencies can be obtained with nano-sized cells which allow for a strong coupling between the photon flux from the sun and the electron flux through the device. In such devices Carnot efficiencies can be reached in the reversible limit where the power output goes to zero. In the situation of maximum power, which is of much greater practical interest, we found that the best efficiencies are remarkably well predicted by the Curzon-Ahlborn result. The presence of non-radiative effects needs to be avoided since it breaks down the strong coupling condition which leads to a drastic decrease in the efficiency.

\begin{table}
\caption{\label{table}The efficiency at maximum power, together with the corresponding values for $E_g$ and $V$. In all cases we take $\Gamma_l=\Gamma_r=\Gamma_s=\Gamma$. }
\begin{ruledtabular}
\begin{tabular}{lrrr}
& $\eta$ (\%) & $E_g$ (eV) & $V$ (V) \\
$\Gamma_{nr}/\Gamma=0.0$ & 77.5 & 0.733 & 0.568 \\
$\Gamma_{nr}/\Gamma=0.1$ & 69.4 & 0.727 & 0.562\\
$\Gamma_{nr}/\Gamma=1.0$ & 36.0 & 0.674 & 0.507 \\
$\Gamma_{nr}/\Gamma=10.0$ & 5.8 & 0.534 & 0.361 \\
\end{tabular}
\end{ruledtabular}
\end{table}

\begin{acknowledgments}
M. E. is supported by the FNRS Belgium (charg\'e de recherches) and by the government of Luxembourg (Bourse de formation recherches). B. C. is a postdoctoral fellow of the FWO - Vlaanderen.
\end{acknowledgments}


\end{document}